\documentclass[aps,prd,groupedaddress,showpacs,nofootinbib,twocolumn]{revtex4}
\usepackage{graphicx,amssymb}

\begin{document}

\title{Constraining $f(R)$ theories with Type Ia Supernovae and Gamma Ray Bursts}

\author{Vincenzo F. Cardone\footnote{{\tt winnyenodrac@gmail.com}}, 
Antonaldo Diaferio\footnote{{\tt diaferio@ph.unito.it}} \&
Stefano Camera\footnote{{\tt camera@ph.unito.it}}}

\affiliation{$^1$Dipartimento di Fisica Generale "Amedeo Avogadro"
and I.N.F.N. - Sezione di Torino, Via Pietro Giuria 1, 10125 - Torino, Italy}

\begin{abstract}

Fourth\,-\,order gravity theories have received much interest in recent years
thanks to their ability to provide an accelerated cosmic expansion in a matter only universe. 
In these theories, the Lagrangian density of the gravitational field has the form 
$R + f(R)$, and the explicit choice of the arbitrary function $f(R)$ must meet 
the local tests of gravity and the constraints from the primordial abundance of 
the light elements.  Two popular classes of $f(R)$ models, which are expected to 
fulfill all the above requirements, have recently been proposed. However, 
neither of these models has ever been quantitatively tested against the available 
astrophysical data.  Here, by combining Type Ia Supernovae and Gamma Ray Bursts, 
we investigate the ability of these models to reproduce the observed
Hubble diagram over the redshift range $(0, 7)$. We find that both models fit very well this 
dataset with the present-day values of the matter density and deceleration parameters which
agree with previous estimates. However, the strong degeneracy among the $f(R)$ parameters
prevents us from putting strong constraints on the values of these parameters; nevertheless,
we can identify the regions of the parameter space that should, in principle, be carefully explored with
future data and dynamical probes in order to discriminate among $f(R)$ theories and
standard dark energy models.

\end{abstract}


\maketitle

\section{Introduction}

It is now widely accepted that the universe is presently undergoing a period
of accelerated expansion. The Hubble diagram of Type Ia Supernovae (SNeIa) 
\cite{SNeIaFirst,SNeIaSec,SNeIaHighZ,SNLS,ESSENCE,D07}, the anisotropy spectrum
of the cosmic microwave background radiation (CMBR) \cite{CMBR,wmap,WMAP5} and the clustering
properties probing the large\,-\,scale structure of the universe \cite{LSS} are concordant
pieces of evidence in favour of this cosmic speed up. In the concordance $\Lambda$CDM cosmological model,
this wide dataset is excellently reproduced \cite{LCDMTest} by assuming a spatially flat universe 
dominated by Cold Dark Matter (CDM) and a cosmological constant $\Lambda$ \cite{Lambda}. 
However, this scenario has two serious drawbacks:
(1) if interpreted as vacuum energy, the $\Lambda$ value is 120
orders of magnitude smaller than what is expected from quantum field 
theory; (2) the coincidence and fine\,-\,tuning
problems do not seem to have a natural explanation. This circumstance 
has motivated the search for alternative explanations. They mostly
rely on scalar fields with a suitable potential which provides a varying $\Lambda$ term or other
dark energy fluids \cite{QuintRev} with exotic properties.

An alternative route towards solving the problem of the accelerated cosmic expansion may be explored by 
looking at the cosmological constant as an additive constant term in the Einstein\,-\,Hilbert
gravity Lagrangian: from the point of view of the field equations, one is not adding any 
new source term to the energy-momentum tensor, but rather it is the
geometrical sector (the left-hand side of the equations) that has been
modified. It is therefore worth wondering whether the accelerated expansion can indeed be obtained
by generalizing this approach, i.e. by adding other non\,-\,linear terms to the action. The search for
a generalization of Einstein General Relativity (GR) actually dates back to just few years after
the seminal Einstein papers (see, e.g., \cite{schmidt} for a historical review), but the relevance
of these corrective terms were considered to be restricted only to the strong gravity regimes. Moreover, 
effective field theory considerations motivated the introduction of small couplings which suppress these 
non\,-\,linear terms in all the other curvature regimes. As a consequence, corrections to GR were considered
only close to the Planck scale, i.e. in the very early\,-\,universe \cite{Star80}, 
and in the attempt to avoid black hole singularities \cite{b-h}. 

The qualitative similarities between  the present\,-\,day acceleration and the early 
universe inflationary expansion renewed the interest in what are now collectively referred 
to as {\it $f(R)$ theories}. Indeed, it became clear soon that an accelerated expansion in 
a matter only universe can be achieved by suitably choosing the functional expression for 
$f(R)$ leading to a non-linear gravity Lagrangian, ${\cal{L}} \propto R + f(R)$. The impressive 
amount of papers \cite{fognoi,fogaltri} investigating different aspects of $f(R)$ theories helps 
understanding that the success in explaining the cosmic acceleration must be balanced
by the constraints provided by the local gravity tests (see, e.g., \cite{FOGrev} for exhaustive reviews 
discussing these issues). Indeed, many models providing an accelerated expansion were later rejected 
because the non-linear terms do not turn off their effect on the Solar System scale and thus lead to
unacceptable contrast with the success of GR in this regime. 

Recently, two models carefully designed  to evade the local gravity tests but still providing 
an accelerated cosmic expansion has been proposed \cite{HS07, Star07}. 
However, neither of these models has yet been quantitatively tested against the
available astrophysical data. Here, in order to to further support (or rule out) 
$f(R)$ theories as alternative candidates to the dark energy hypothesis, we search for the 
appropriate set of parameters of these models to verify that 
the acceleration, that they qualitatively predict, actually quantitevely agree
with the data.  

The scheme of the paper is as follows. In \S\,II, we will
briefly remind the basics of $f(R)$ theories writing down the
equations needed to determine the background evolution of the
universe. \S\,III discusses the theoretical constraints driving 
the choice of a functional expression for $f(R)$ thus motivating the 
adoption of the two popular models to be tested.  The data used
and the statistical method used to constrain the model parameters are
presented in \S\,IV, while the results of the analysis are 
commented upon in \S\,V. Conclusions and future perspectives are
finally given in \S\,VI. 

\section{$f(R)$ cosmology}

The idea of modifying the gravity sector of the Einsteinian
General Relativity (GR) dates back to the Starobinsky \cite{Star80}
attempts to get an inflationary expansion of the early universe 
without the need of any scalar field. This can be accomplished 
by generalizing the Einstein\,-\,Hilbert action as\,:
\begin{equation}
S = \int{d^4x \sqrt{-g} \left [ \frac{R + f(R)}{2 \kappa^2}
+ {\cal{L}}_M \right ]}
\label{eq: fogaction}
\end{equation}
where $R$ is the scalar curvature, $\kappa^2 = 8 \pi G$ (we
use units where the light speed $c=1$), 
${\cal{L}}_M$ is the standard matter Lagrangian, and $f(R)$ is an 
analytical function expressing the deviation from the Einstein GR. 
For $f(R) = -2 \Lambda$, we get back the concordance $\Lambda$CDM model,
while nontrivial dynamics is obtained for other choices. 

Under the metric approach to $f(R)$ gravity, the field equations
are obtained by varying the action (\ref{eq: fogaction}) with 
respect to the metric only. We obtain\,:
\begin{equation}
G_{\alpha \beta} + f_R R_{\alpha \beta}
- \left ( \frac{1}{2} f - \Box f_R \right ) g_{\alpha \beta}
- \nabla_{\alpha} \nabla_{\beta} f_R = \kappa^2 T_{\alpha \beta}
\label{eq: fogfield}
\end{equation}
with $G_{\alpha \beta} = R_{\alpha \beta} - (R/2) g_{\alpha \beta}$
and $T_{\alpha \beta}$ the usual Einstein and source stress\,-\,energy
tensor. Hereafter the subscript $R$ will denote differentiation
with respect to $R$. We also assume a spatially  flat Robertson\,-\,Walker metric,
with scale factor $a$; in this case, the scalar curvature reads

\begin{equation}
R = 12 H^2 + 6 H H^{\prime}
\label{eq: defR}
\end{equation}
where $H = \dot{a}/a$ is the Hubble parameter and the dot indicates
the time derivative; by using the more
convenient variable $\eta = \ln{a}$, we denote
the derivative with respect to $\eta$ with a prime. With
these definitions, Eqs.(\ref{eq: fogfield}) lead to the 
modified Friedmann equation\
\begin{equation}
H^2 = f_R \left ( H H^{\prime} + H^2 \right ) 
+ \frac{1}{6} f + H^2 f_{RR} R^{\prime} = \frac{\kappa^2}{3}
\rho_M \ ,
\label{eq: fogfried}
\end{equation}
where we have assumed that dust matter with enery density 
$\rho_M$ and pressure $p_M = 0$ is the only fluid filling 
the universe. To solve Eqs.(\ref{eq: defR}) and (\ref{eq: fogfried}),
we follow \cite{HS07} and introduce the dimensionless variables\,:
\begin{equation}
y_H = \frac{H^2}{m^2} - a^{-3} \ \ , \ \ 
y_R = \frac{R}{m^2} - 3 a^{-3} \ \ ,
\label{eq: defyvar}
\end{equation}
with 

\begin{equation}
m^2 = \frac{\kappa^2 \rho_M(\eta = 0)}{3} 
\simeq (8315 \ {\rm Mpc})^{-2} \ \left ( 
\frac{\Omega_M h^2}{0.13} \right )
\label{eq: defm}
\end{equation}
a convenient curvature scale depending on the present day 
values of the matter density parameter $\Omega_M$ and 
the Hubble constant $h = H_0/(100 \ {\rm km/s/Mpc})$. 
The background evolution of the universe is determined
by the following set of coupled ordinary differential 
equations\,:
\begin{equation}
y_H^{\prime} = \frac{1}{3} y_R - 4 y_H \ , 
\label{eq: yhpeq}
\end{equation}
\begin{eqnarray}
y_R^{\prime} & = & 9 a^{-3} - \frac{1}{m^2 f_{RR} 
\left ( y_H + a^{-3} \right )} \nonumber \\
~ & \times & \left [ y_H - \left ( \frac{1}{6} y_R
- y_H - \frac{a^{-3}}{2} \right ) f_R + \frac{f}{6 m^2} \right ] \ .
\label{eq: yrpeq}
\end{eqnarray}
It is worth noting that the above system of equations 
of first order in $(y_H, y_R)$ is equivalent
to a single second order equation in $y_R$. If we look at the 
definition of $y_R$, we see that the full system is equivalent to
a single fourth-order non-linear differential equation for the
scale factor $a(t)$: this property explains why $f(R)$ theories are
also commonly referred to as fourth-order gravity theories.

In order to integrate the system, we need to set the boundary 
conditions. In \cite{HS07}, the authors set them 
at redshifts $z\to \infty$, by requiring that $f(R)$ tends to a constant
at this epoch and by considering the detailed balance of the 
perturbative corrections to $R = \kappa^2 \rho_M$. However, 
here we are interested in fitting the model to data that probe
the range $z<10$; therefore, it is more interesting to set the
present day values of $(y_H, y_R)$ and integrate the equations back in time 
up to $a \simeq 0.001$ $(\eta \simeq -7)$. By setting, as usual,
$a_0 = 1$ and remembering that 
\begin{equation}
R_0 = 6 H_0^2 (1 - q_0)
\label{eq: rzqz}
\end{equation}
with $q = -\ddot{a} a/\dot{a}^2$ the deceleration parameter,
we get:
\begin{equation}
y_H(0) = H_0^2/m^2 - 1 \ \ , \ \ 
y_R(0) = 6 (H_0^2/m^2) (1 - q_0) - 3 \ \ .
\label{eq: incond}
\end{equation}
It is worth noting that, because of Eq.(\ref{eq: defm}),
the initial conditions (\ref{eq: incond}) are determined
by the values of the three parameters $(\Omega_M, h, q_0)$. 
This is again consistent with $f(R)$ models being fourth-order 
theories; thus, we require three initial conditions to
determine the evolution of the scale factor for any
given $f(R)$. Had we written down 
a single equation for $a(t)$, we should have set the 
present\,-\,day values of the first three time derivatives of
$a(t)$ which include the jerk parameter 
$j = (d^3a/dt^3) H^{-3}$. In contrast, the introduction
of the $(y_H, y_R)$ auxiliary functions and of the curvature scale
$m$ enables us to replace the quite uncertain jerk parameter
with the more manageable matter density $\Omega_M$.

\section{$f(R)$ models}

A key role in fourth\,-\,order gravity is obviosuly played by
the functional expression of $f(R)$. In principle, such a choice
is fully arbitrary unless one has a theoretical motivation 
leading to a unique expression for $f(R)$. Actually, things are 
not so easy. Indeed, the modification of GR introduces deviations not only
on the cosmological scales, but at all the scales where gravitational
phenomena can be tested. In particular, it has proven to be quite 
difficult for a large class of $f(R)$ theories to evade the constraints
on the Solar System scale (see, e.g., \cite{nosolsys} and refs therein). 
As Chiba \cite{chiba} has shown, the main difficulty arises from $f(R)$
models introducing a new scalar degree of freedom with the same coupling
to matter as gravity. As a consequence, it appears a long range fifth force that
violates the constraints on the PPN parameters. Although the derivation
of the PPN parameters has been questioned \cite{CST07}, significant deviations 
from the GR metric around the Sun seem to be confirmed because of a decoupling
of the scalar curvature from the local density. As a possible way out, one can
invoke a chamaleon effect \cite{cham} to reassociate high density with high 
curvature so that the scalar degree of freedom becomes very massive and the
fifth force escapes any detection. To this aim, $f(R)$ should be tailored in
such a way to give rise to a mass squared term which is large and positive 
in high curvature environments \cite{frcham}. It is worth stressing that 
this same condition is also required if we want to recover 
GR in the early universe \cite{frearly} and obtain the usual matter
dominated era. Since $R \rightarrow \infty$ in this limit, we 
expect that $f(R)$ tends to a small constant in order to make its effect 
negligible with respect to the GR term. On the other hand, in the late universe, 
we expect to mimic the same evolutionary history of the $\Lambda$CDM because 
this model agrees with the data; therefore $f(R)$ should 
reduce again to a small constant, but it should again tend to zero in the limit of a vanishing $R$ 
to agree with the observational fact that $\Lambda$ takes a very low value.
Summarizing, one has to look for a functional expression satisfying 
the following constraints:
\begin{equation}
\left \{
\begin{array}{l}
\displaystyle{\lim_{R \rightarrow 0}{f(R)} = 0} \\
~ \\
\displaystyle{\lim_{R \rightarrow \infty}{f(R)} = {\rm const}} \\
~ \\
\displaystyle{f_R(R)|_{R \gg  m^2} = df(R)/dR|_{R \gg m^2} > 0} \\
~ \\
\displaystyle{f_{RR}(R)|_{R \gg m^2} = d^2f(R)/dR^2|_{R \gg m^2} > 0}
\end{array}
\right . \ ,
\label{eq: frconstr}
\end{equation}
where $m^2$ is a typical curvature scale and the last condition \cite{frstab}
ensures that, in the limit $R \gg m^2$, the solution is stable at high curvature.

Among the possible choices left out by the above conditions, we will consider here
two classes of $f(R)$ models. For the first one, we follow \cite{HS07} and set:
\begin{equation}
f(R) = - m^2 \frac{c_1 (R/m^2)^n}{1 + c_2 (R/m^2)^n} 
\label{eq: frhs}
\end{equation}
with $m$ given by (\ref{eq: defm}), and $(n, c_1, c_2)$ are positive
dimensionless constants. We will refer to this choice as the 
Hu \& Sawicki (hereafter, HS) model by the name of the authors
who first suggested this expression. It is easy to check that all
the constraints (\ref{eq: frconstr}) are easily passed by the
HS model. In particular, we note that, since $f(0) = 0$, there is
no actual cosmological constant in the model, but
\begin{displaymath}
\lim_{m^2/R \rightarrow 0}{f(R)} \simeq 
- \frac{c_1}{c_2} m^2 + \frac{c_1}{c_2^2} m^2 \left (
\frac{m^2}{R} \right )^n
\end{displaymath}
so that, when $c_1/c_2^2 \rightarrow 0$ at fixed 
$c_1/c_2$, we recover an effective cosmological constant in 
high curvature $(m^2/R \rightarrow 0)$ environments.  

Another possibility to satisfy all the constraints (\ref{eq: 
frconstr}) is offered by the Starobinsky proposal \cite{Star07}\,:
\begin{equation}
f(R) = \lambda R_{\star} \left [ \left (
1 + \frac{R^2}{R_{\star}^2} \right )^{-n} - 1 \right ]
\label{eq: frst}
\end{equation} 
with $R_{\star}$ a scaling curvature parameter and $(\lambda, n)$
two positive constants. We will refer to this class of $f(R)$
theories as the Starobinsky (St) model. Note that, even in this
case, $f(0) = 0$ so that no actual cosmological constant is present. 
Nevertheless, an effective one is recovered in the high curvature
regime as can be seen from $f(R \gg R_{\star}) \sim -2 \Lambda_{\infty}$
with $\Lambda_{\infty} = \lambda R_{\star}/2$.

As a general remark, we note that the HS and St models are quite similar
at both very low and very high redshifts since they are
both built up by imposing the same constraints on $f(R)$. Moreover, they 
both aim at mimicking the successful $\Lambda$CDM scenario in the late 
and early universe. Put in other words, the HS and St models both reduces
to the GR\,+\,$\Lambda$ case in the limits of very high and very low curvature.
What makes them different is the way the two extreme cases are connected, i.e. 
how the universe evolves from the present day $\Lambda$ dominated phase to the 
early matter epoch. 

For completeness, we finally remind the reader that the two models we 
are considering here are not the only viable ones; other possible 
examples are given in \cite{AB07,SO07}. It is, moreover, possible to work 
out $f(R)$ models which can also provide an inflationary expansion in the 
very early universe \cite{SOinfl}. However, all these other cases share 
many similarities with the HS and St models so that we are confident that
exploring just two classes of fourth-order gravity theories should provide us
with some general conclusions on their viability. 

\section{Constraining the models}

Any model that aims to describe the evolution of the
universe must be able to reproduce what is indeed observed. 
Matching the model with observations is also a powerful 
tool to constrain its parameters and allows to estimate some further 
quantities of interest. 

As mentioned above, the HS and St models are carefully
designed to give an effective cosmological constant in the
late universe. Moreover, both theories are assigned by 
a three parameters function, and one can anticipate that the
considerable freedom allowed by the degeneracy among the parameters
makes it easy to find some combinations that lead to quite 
similar $H(z)$ in the low $z$ regime. Fitting 
to SNeIa data can only tell us whether the models are viable over
the redshift range $(0, 1.5)$. We expect that this is indeed the case 
because the $f(R)$ functions have been tailored to do so. 
What is not garanteed is that 
the HS and St $f(R)$ models can describe the background
expansion up to higher redshifts $z$.
To probe this regime, we will 
use the recently derived Hubble diagram of Gamma Ray Bursts (GRBs). 

In the Bayesian approach to model testing, we explore the parameter 
space through the likelihood function\,:
\begin{eqnarray}
{\cal{L}}({\bf p}) & \propto  & {\cal{L}}_{SNeIa}({\bf p}) \ 
{\cal{L}}_{GRB}({\bf p}) \times \nonumber \\
~ & \times &  \exp{\left [ - \frac{1}{2}
\left ( \frac{\omega_M^{\rm obs} - \omega_M^{\rm th}}{\sigma_M}  \right )^2
\right ]} \times \nonumber \\
~ & \times & \exp{\left [ - \frac{1}{2} 
\left ( \frac{h_{HST} - h}{\sigma_{HST}} \right )^2 \right ]}
\label{eq: deflike}
\end{eqnarray}
where {\bf p} denotes the set of model parameters. Before discussing in
detail the term related to the SNeIa and GRB data, we concentrate on
the two Gaussian priors. The former takes into account the 
constraints on the physical matter density $\omega_M = \Omega_M h^2$ with
\begin{displaymath}
\omega_M^{\rm obs} \pm \sigma_M = 0.137 \pm 0.004
\end{displaymath}
as inferred from WMAP5 data \cite{WMAP5}. One can wonder whether such an 
estimate may be used as a constraint on the $f(R)$ models since it has been
obtained by fitting the CMBR spectrum assuming the validity of GR. However,
what is really needed for this estimate to be model independent is not that 
GR holds along all the evolutionary history, but that the gravity Lagrangian
reduces to the Einstein\,-\,Hilbert one at the last scattering. Since, for 
both HS and St models, $f(R)/R \rightarrow 0$ for $z \simeq 1000$, we can
safely use the WMAP5 $\omega_M$ value as a constraint. 

The Gaussian prior on $h$ in Eq.(\ref{eq: deflike}) stems from the results 
of the HST Project \cite{hstkey} which has estimated the Hubble constant 
$H_0$ using a well calibrated set of local distance scale. Averaging over the 
different methods, the survey finally gives\,:
\begin{displaymath}
h_{HST} \pm \sigma_{HST} = 0.72 \pm 0.08
\end{displaymath}
as a cosmological model independent constraint\footnote{While this work was 
near completion, the SHOES collaboration \cite{shoes} has released a more precise
estimate as $h = 0.742 \pm 0.036$ in agreement with our adopted value.}. 

The main two terms in the likelihood (\ref{eq: deflike}) are both
related to the Hubble diagram, the first one being, in particular, 
connected with SNeIa. These latter data have provided the first 
piece of evidence for the cosmic speed up and are still a sort of {\it ground-zero} 
test that every cosmological model has to pass to be considered acceptable. 
To check this, one relies on the predicted distance modulus\,:
\begin{equation}
\mu_{th}(z, {\bf p}) = 25 + 5 \log{\left [ \frac{c}{H_0} (1 + z) r(z, {\bf p}) \right ]}
\label{eq: defmuth}
\end{equation}
with $r(z)$ the dimensionless comoving distance\,:
\begin{equation}
r(z, {\bf p}) = \int_{0}^{z}{\frac{dz'}{E(z', {\bf p})}} \ .
\label{eq: defrz}
\end{equation}
The likelihood function is then defined as
\begin{eqnarray}
{\cal{L}}_{SNeIa}({\bf p}) & = & 
\frac{1}{(2 \pi)^{{\cal{N}}_{SNeIa}/2} |C_{SNeIa}^{-1}|^{1/2}}
\nonumber \\ ~ & \times &
\exp{\left ( - 
\frac{\Delta \mu \cdot C_{SNeIa}^{-1} \cdot \Delta \mu^T}{2} 
\right )} \ ,
\label{eq: deflikesneia}
\end{eqnarray}
where ${\cal{N}}_{SNeIa}$ is the total number of SNeIa used, 
$\Delta \mu$ is a ${\cal{N}}_{SNeIa}$\,-\,dimensional vector with 
the values of $\mu_{\rm obs}(z_i) - \mu_{\rm th}(z_i)$ and $C_{SNeIa}$
is the ${\cal{N}}_{SNeIa} \times {\cal{N}}_{SNeIa}$ covariance matrix 
of the SNeIa data. Note that, if we neglect the correlation induced by 
systematic errors\footnote{Actually, it has been shown \cite{K08} that
neglecting systematic errors does not shift the central values, but only 
weakens the constraints. Although such a result has been obtained using 
standard dark energy models, we are confident that this is also the case 
for our $f(R)$ theories.}, as we do here, $C_{SNeIa}$ is a diagonal matrix and 
Eq.(\ref{eq: deflikesneia}) simplifies to\,:
\begin{equation}
{\cal{L}}_{SNeIa}({\bf p}) \propto \exp{[-\chi_{SNeIa}^2({\bf p})/2]}
\label{eq: likesneiasimple}
\end{equation}
with
\begin{equation}
\chi_{SNeIa}^2({\bf p}) = 
\sum_{i = 1}^{{\cal{N}}_{SNeIa}}{\left [ 
\frac{\mu_{obs}(z_i) - \mu_{th}(z_i)}{\sigma_i} \right ]^2}
\label{eq: defchisneia}
\end{equation}
with $\sigma_i$ the error on the observed distance modulus 
$\mu_{obs}(z_i)$ for the i\,-\,th object at redshift $z_i$. As 
input data, we use the Union SNeIa sample assembled in \cite{K08} by
reanalysing with the same pipeline both the recent SNeIa SNLS \cite{SNLS}
and ESSENCE \cite{ESSENCE} samples and older nearby and high redshift 
\cite{SNeIaHighZ} datasets.

Although quite useful in probing the accelerated expansion, SNeIa are 
limited to $z \sim 1.5$. As a consequence, one 
has to resort to a different distance indicator to push the Hubble diagram 
to higher redshift and probe the (supposedly) matter dominated era. Thanks to 
the enormous energy release that makes them visibile up to $z \sim 6.6$, GRBs 
stand out as ideal candidates to this scope. The discovery of 2D correlations 
between their properties have opened the way towards making GRBs 
standard candles similarly to SNeIa \cite{CCD09}. 
As a result, Schaefer \cite{Schaefer} have
provided the first GRBs Hubble diagram containing 69 objects with $\mu_{\rm obs}(z)$
estimated by averaging over 5 different 2D correlations. We use here the updated GRBs 
Hubble diagram recently presented in \cite{CCD09} based on a model-independent
recalibration of the same 2D correlations used by Schaefer. 

Since there is no correlation among the errors of different GRBs, 
the likelihood function now simply reads\,:
\begin{equation}
{\cal{L}}_{GRB}({\bf p}) \propto \exp{[-\chi_{GRB}^2({\bf p})/2]}
\label{eq: likegrbsimple}
\end{equation}
with
\begin{equation}
\chi_{GRB}^2({\bf p}) = 
\sum_{i = 1}^{{\cal{N}}_{GRB}}{\left [ 
\frac{\mu_{obs}(z_i) - \mu_{th}(z_i)}{\sqrt{\sigma_i^2 + \sigma_{GRB}^2}} \right ]^2}
\label{eq: defchigrb}
\end{equation}
where $\sigma_{GRB}$ takes care of the intrinsic scatter inherited from
the scatter of GRBs around the 2D correlations used to derive the 
individual distance moduli.\footnote{Note that a similar term is also 
present for SNeIa, but it is estimated to be $\sigma_{int} = 0.15$ and 
yet included into the error $\sigma_i$ provided in the Union dataset.}

\section{Results}

The $f(R)$ functions for the HS and St models in Eqs.(\ref{eq: frhs}) 
and (\ref{eq: frst}) depend on three parameters, while other three
parameters are needed to set the initial conditions (\ref{eq: incond}). 
By adding the GRBs intrinsic scatter $\sigma_{GRB}$, we end up with a 
seven dimensional parameter space to be explored. To do this efficiently,
we use a Markov Chain Monte Carlo code which maximizes the likelihood 
(\ref{eq: deflike}) along a chain with $\sim 400,000$ points. The 
constraints on the parameters are then obtained by cutting out 
the first $30\%$ of the chain; we thus skip the burn-in phase and thin 
the chain to reduce spurious correlations. 

Before discussing the results, we warn the reader that, in the context 
of Bayesian statistics, the best fit model, i.e, the set of parameters 
${\bf p}$ maximizing ${\cal{L}}({\bf p})$, represents the most plausible 
model in an Occam's razor sense given the data at hand. However, in a
Bayesian context, the best fit parameters individually do not necessarily
have to be probable, but rather they must have a high joint probability density
that might occupy only a small volume of the parameter space.
This situation can arise if the best fit solution does not lie in the bulk
of the posterior probability distribution. Such a situation may often occur  
when the posterior is non-symmetric in a high dimensional space so that the volume
can dramatically increase with the distance from the best fit solution. In this
case, the best fit solution for each parameter could easily lie outside the bulk
of the individual posterior distribution for $p_i$ obtained by marginalizing over the
other parameters. This is indeed what happens for our models so that we have
preferred to remind the reader that this somewhat counterintuitive outcome is 
not a problem, but rather a common feature in statistics in multi-dimensional spaces.

\subsection{The HS model}

The HS $f(R)$ functional expression depends on three dimensionless 
positively defined parameters. While it is reasonable to expect that
$n$ is not a large quantity, nothing can be said a priori on the order
of magnitude of $(c_1, c_2)$. Indeed, should both of them be very large, 
then eq.~(\ref{eq: frhs}) reduces to $f(R, c_1 \gg 1, c_2 \gg 1) \sim 
-m^2 (c_1/c_2)$ so that this case provide an effective cosmological constant.
In contrast, should $c_2$ be very small, we get $f(R) \sim R^n$ thus 
recovering power\,-\,law corrections to the Einstein\,-\,Hilbert Lagrangian
which are known to provide accelerated expansions for $n \sim 2$. We must 
therefore explore a huge range for $(c_1, c_2)$; we thus skip to 
logarithmic units and get constraints on $(\log{c_1}, \log{c_2})$ by
running the MCMC code.

The best fit model turns out to be\,:

\begin{figure}
\centering \resizebox{8.5cm}{!}{\includegraphics{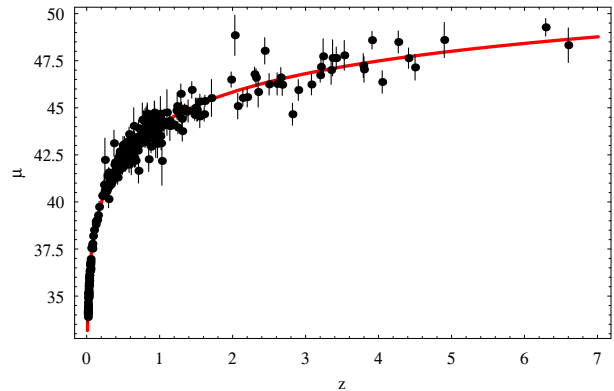}}
\caption{Best fit HS model superimposed to the data.}
\label{fig: bfhsplot}
\end{figure}

\begin{displaymath}
\Omega_M = 0.282 \ , \ h = 0.703 \ , \ q_0 = -0.67 \ ,
\end{displaymath}
\begin{displaymath}
n = 4.26 \ , \ \log{c_1} = -0.53 \ , \ \log{c_2} = -8.39 \ ,
\end{displaymath}
with a best fit GRB intrinsic scatter $\sigma_{GRB} = 0.41$. The 
overall quality of the fit may be seen by looking at Fig.\,\ref{fig: bfhsplot}
and quantified by considering the following estimators\,:

\begin{displaymath}
\chi^2_{SNeIa}/d.o.f. = 1.03 \ , \ 
\chi^2_{GRB}/d.o.f. = 1.17 \ , \ 
\omega_M = 0.139 
\end{displaymath}
so that we can safely conclude that the HS model is in very
good agreement with the data. A cautionary note is in order here 
concerning the $\chi^2$ values reported above. The MCMC code maximizes 
the likelihood (\ref{eq: deflike}) which is strictly not the same as 
minimizing either $\chi^2_{SNeIa}$ or $\chi^2_{GRB}$. Moreover, from 
a statistical point of view, the significance level of the above reduced
$\chi^2$ values can not be estimated from the usual tables since these 
standard results do not take into account systematic errors or intrinsic 
scatter. As such, both $\chi^2_{SNeIa}/d.o.f.$ and $\chi^2_{GRB}/d.o.f.$ 
must be considered only as a useful tool to quantify the agreement with the data,
but should not be overrated. 

The constraints on the single parameters are summarized in Table I where we give the mean and
median values and the $68\%$ and $95\%$ confidence limits. While the 
best fit solution for $(\Omega_M, h, q_0)$ is quite close
to the median values, this is not the case for $(n, \log{c_1},
\log{c_2})$, with the best fit solution for $n$ lying marginally
outside the $95\%$ CL. However, according to the Bayesian philosophy, 
one must take the constraints in Table I as the final outcome of the
likelihood analysis. It is, however, worth noting that setting all the 
parameters to their median values gives\,:

\begin{displaymath}
\chi^2_{SNeIa}/d.o.f. = 1.04 \ , \ 
\chi^2_{GRB}/d.o.f. = 1.45 \ , \ 
\omega_M = 0.139 
\end{displaymath}
so that the quality of the fit is still not unreasonably bad. 

\begin{table}[t]
\begin{center}
\begin{tabular}{cccccc}
\hline
$x$ & $x_{BF}$ & $\langle x \rangle$ & $x_{med}$ & $68\%$ CL & $95\%$ CL \\
\hline \hline
~ & ~ & ~ & ~ & ~ & ~ \\
$\Omega_M$ & 0.282 & 0.282 & 0.282 & (0.268, 0.296) & (0.256, 0.309) \\
~ & ~ & ~ & ~ & ~ & ~ \\
$h$ & 0.703 & 0.700 & 0.700 & (0.692, 0.708) & (0.685, 0.716) \\
~ & ~ & ~ & ~ & ~ & ~ \\
$q_0$ & -0.67 & -0.63 & -0.62 & (-0.70, -0.55) & (-0.88, -0.45) \\
~ & ~ & ~ & ~ & ~ & ~ \\
$n$ & 4.26 & 2.52 & 2.36 & (1.84, 3.24) & (1.44, 4.21) \\
~ & ~ & ~ & ~ & ~ & ~ \\
$\log{c_1}$ & -0.53 & 1.15 & 1.00 & (-1.07, 2.96) & (-2.31, 7.12) \\
~ & ~ & ~ & ~ & ~ & ~ \\
$\log{c_2}$ & -8.39 & -7.64 & -7.65 & (-9.23, -6.11) & (-9.90, -5.27) \\
~ & ~ & ~ & ~ & ~ & ~ \\
$\sigma_{GRB}$ & 0.41 & 0.29 & 0.29 & (0.17, 0.41) & (0.03, 0.50) \\
~ & ~ & ~ & ~ & ~ & ~ \\
\hline
\end{tabular}
\end{center}
\caption{Constraints on the HS model parameters.}
\end{table}

In order to further investigate the viability of the model, we 
can compare the constraints on $(\Omega_M, q_0)$ with
other results in the literature.\footnote{We no longer consider 
anymore the Hubble constant $h$ because this is typically 
marginalized over when fitting Hubble\,-\,diagram data because
of the degeneracy with the (unknown) SN absolute magnitude. Such 
a degeneration is partially broken here thanks to the use 
of two different distance indicators and the Gaussian prior 
from the HST Key Project; however, we prefer not to discuss 
the corresponding constraints because other model independent estimates
in the literature (coming from, e.g., time delays in multiply lensed
quasars) are affected by too large uncertainties.} However,
the matter density parameter $\Omega_M$ is always estimated by fitting
a given model to a certain set of data so that a straightforward 
comparison may be biased by the different theory we are considering. 
Therefore, we simply note that the values in Table I are in very 
good agreement with typical estimates \cite{D07,K08,WMAP5} 
from previous analyses of comparable datasets. 

Something more interesting can be said on the deceleration 
parameter $q_0$. Indeed, the problem of model dependent
estimates can now be avoided by resorting to cosmographic 
analyses based only on the Taylor expansion of the scale factor. By using this approach, 
Catto\"en and Visser \cite{CV08} have found values between 
$q_0 = -0.48 \pm 0.17$ and  $q_0 = -0.75 \pm 0.17$, depending on the 
details of the  method used to fit the SNLS dataset. A similar analysis, 
but using a different and smaller GRBs sample, enabled Capozziello 
and Izzo \cite{CI08} to find values between $q_0 = -0.94 \pm 0.30$
and $q_0 = -0.39 \pm 0.11$, still in agreement with our
estimates. A different approach has been instead adopted 
by Elgar\o y and Multam\"aki \cite{EM06} which advocate a model-independent 
parametrization of $q(z)$. Depending on the SNeIa sample used and 
the parametrization adopted, their best fit values for $q_0$ 
range between $-0.29$ and $-1.1$ still in good agreement with our
constraints in Table I.

\begin{figure}
\centering \resizebox{8.5cm}{!}{\includegraphics{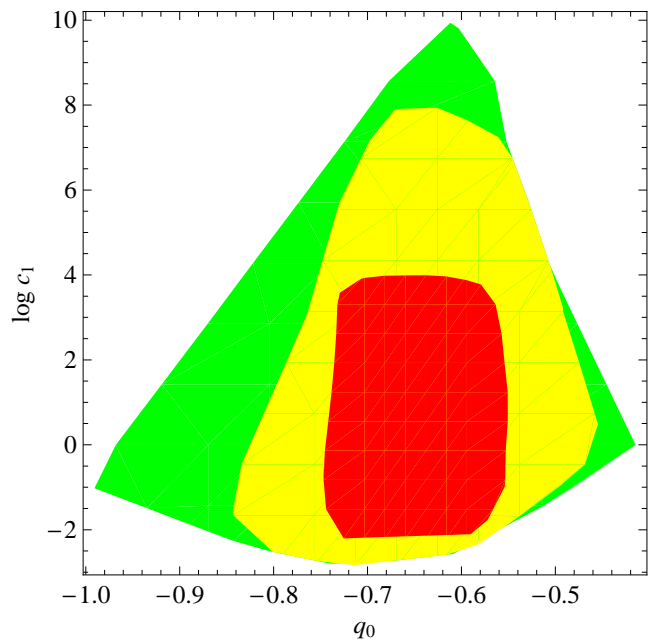}}
\caption{Likelihood countours in the plane $(q_0, \log{c_1})$. Red, yellow 
and green shaded regions refer to the $68$, $95$, $99\%$ confidence regions 
respectively.}
\label{fig: qzc1cont}
\end{figure}

While the value of $q_0$ tells us that the model is nowadays undergoing
accelerated expansion, it is also important to check that this cosmic
speed up ends well before the epoch of structure formation. To
do so, for each point along the Markov chain, we compute the deceleration
parameter $q(z)$ and solve the equation $q(z_T) = 0$ with $z_T$ usually referred
to as the transition redshift. We then make a histogram of the $z_T$ values 
and use this as a proxy for its probability distribution so finally estimating\,:

\begin{displaymath}
\langle z_T \rangle = 0.52 \ \ , \ \ 
z_{T,med} = 0.51 \ \ ,
\end{displaymath}
\begin{displaymath}
{\rm 68\% \ CL \ :} \ \ (0.44, 0.58) \ \ , \ \ 
{\rm 95\% \ CL \ :} \ \ (0.37, 0.70) \ \ .
\end{displaymath}
These constraints may be compared with previous results. 
For instance, by using the Gold SNeIa sample  and linearly expanding $q(z)$, 
Riess et al. \cite{SNeIaHighZ} found $z_T = 0.46 \pm 0.13$ 
in agreement with the updated result $z_T = 0.49_{-0.07}^{+0.14}$ 
obtained by Cuhna \cite{Cu09} which use the Union sample. Although there 
is a very good agreement, we nevertheless warn the reader that
the estimate of $z_T$ is strongly model dependent. To be conservative,
we may therefore only conclude that the transition to a decelerated 
epoch in the HS model takes place early enough not to conflict with
the growth of structures and galaxy formation. 

Although no previous analysis of the HS model has been performed,
it is worth considering the constraints on the $f(R)$ parameters 
$(n, \log{c_1}, \log{c_2})$. The first striking result is the 
very high value of $\log{c_1}$ with $\log{c_2}$ being, in contrast,
very low. Unless we are in the early universe when $R/m^2$ also 
takes a very high value, the bulk of the models with $(\log{c_1}, 
\log{c_2})$ in Table I, the HS gravity Lagrangian may be approximated as\,:
\begin{displaymath}
R + f(R) \sim R - m^2 c_1 \left ( \frac{R}{m^2} \right )^n 
\end{displaymath}
with the low value of $m$ compensating for the large $c_1$
so that the two terms $R$ and $R^n$ are comparable. From Table I,
we also get $n \sim 2.5$ so that the likelihood analysis is selecting
a subclass of the HS model that behaves as if the gravity Lagrangian were 
$R + (R/m^2)^{2.5}$ in the late universe. 

In a sense, such a result could be anticipated by remembering the history of fourth order 
gravity theories. Indeed, as quoted above, the first successfull $f(R)$
model was proposed by Starobinsky in the '80s to give an accelerated
expansion by adding  a quadratic term to the Einstein\,-\,Hilbert Lagrangian, 
i.e. setting $f(R) \propto R^2$. Indeed, the HS $f(R)$
reduces to the Starobinsky inflationary model for $n = 2$ and $c_2 = 0$ with
$c_1$ weighting the relative importance of the two terms $R$ and $R^2$. Indeed, 
by looking at the likelihood contours in the $(q_0, \log{c_1})$ plane shown in 
Fig.\,\ref{fig: qzc1cont}, one can see that, for a given $\log{c_1}$ value, there are
two $q_0$ solutions, depending on the $n$ being larger or smaller than 2. For $n \ge 2$,
the solution with the smaller $q_0$ provides a better fit to the data. In this $n$ regime, 
one gets that the more weight one gives to the corrective term, the more accelerated is 
the present day expansion. It is therefore not surprising that our likelihood analysis 
has indeed converged not too far from the inflationary $f \sim R^2$ model. 

We stress, however, that, even with $c_1 \gg 1$ and $c_2 \ll 1$,
the HS model remains fundamentally different from the Starobinsky inflationary model. 
Indeed, the accelerated expansion is now obtained at the present day rather than
in the early universe. For very high $z$, $R/m^2$ has become so large that the 
term $c_2 (R/m^2)^n$ in the denominator of Eq.~(\ref{eq: frhs}) is dominant and
we recover the limit $f(R) \sim -m^2 (c_1/c_2)$, i.e. an effective cosmological 
constant. Moreover, it can be shown that, although large, this constant 
term is nevertheless negligible with respect to $R$ so that we fully
recover the GR Lagrangian and we do not thus violate the constraints from BBN and the 
abundance of primordial light elements.

\subsection{The St model}

As for the HS case, the $f(R)$ expression (\ref{eq: frst}) 
of the St model still is a function with three parameters, namely 
$(n, \lambda, R_{\star})$. It is, however, convenient to 
reparameterize the model in terms of dimensionless quantities. 
To this aim, we first define\,:
\begin{equation}
\varepsilon = R_{\star}/R_0 \ \ , \ \ 
\tilde{\Lambda}_{eff} = - \lambda R_{\star}/6 H_0^2 \ \ ,
\label{eq: defnewparst}
\end{equation}
with $R_0$ the present day scalar curvature so that 
Eq.(\ref{eq: frst}) becomes\,:
\begin{equation}
f(R) = - 6 H_0^2 \tilde{\Lambda}_{eff} \left [
\left ( 1 + \frac{R^2}{\varepsilon^2 R_0^2} \right )^{-n} - 1 \right ] \ .
\label{eq: frhsnew}
\end{equation}
Concerning the values of the two parameters $(\varepsilon, \tilde{\Lambda}_{eff})$,
we can only have some hints on what their order of magnitude can be. To this aim,
we first note that $R_{\star}/R_0 = R(z_{\star})/R_0$, with $z_{\star}$ an unknown
reference redshift. A plot of $R(z)/R_0$ for a fiducial $\Lambda$CDM model shows that
this ratio can easily take very large values even at intermediate redshift. Needless to
say, the St model is not the $\Lambda$CDM one, but we can anticipate that, 
in order to fit the data, its expansion rate will not differ much 
so that $R(z)/R_0$ will take similar values. As a result, we therefore
end up with $\varepsilon = R_{\star}/R_0$ taking high values so that we skip
to $\log{\varepsilon}$ as a parameter to be constrained by the MCMC analysis. To 
get an idea of the range for $\tilde{\Lambda}_{eff}$, we note that, in the early 
universe, the Lagrangian reduces to\,:

\begin{figure}
\centering \resizebox{8.5cm}{!}{\includegraphics{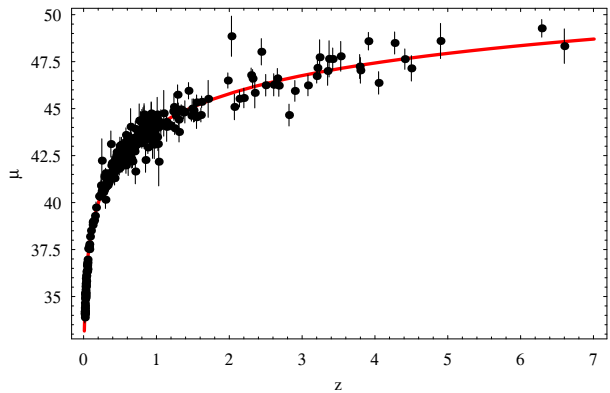}}
\caption{Best fit St model superimposed to the data.}
\label{fig: bfstplot}
\end{figure}

\begin{displaymath}
R + f(R) \sim R - 6 H_0^2 \tilde{\Lambda}_{eff} = R - 2 \Lambda_{\infty}
\end{displaymath}
so that $\tilde{\Lambda}_{eff} = \Lambda_{\infty}/3H_0^2$. Since we want
to recover the GR in this limit, we just have to ask that $R/\Lambda_{\infty} \gg 1$.
However, this ratio can easily be very small even if $\Lambda_{\infty}$ is quite
high so that we end up with $\tilde{\Lambda}_{eff}$ varying over a wide range. 
Therefore, we skip again to logarithmic units taking $\log{\tilde{\Lambda}_{eff}}$ as 
a model parameter.

Running the MCMC gives us as best fit parameters\,:

\begin{displaymath}
\Omega_M = 0.283 \ , \ h = 0.704 \ , \ q_0 = -0.78 \ ,
\end{displaymath}
\begin{displaymath}
n = 1.44 \ , \ \log{\varepsilon} = 1.56 \ , \ 
\log{\tilde{\Lambda}_{eff}} = 2.86 \ ,
\end{displaymath}
with a best fit GRB intrinsic scatter $\sigma_{GRB} = 0.38$. 
The values of the reduced $\chi^2$ and of $\omega_M$

\begin{table}[t]
\begin{center}
\begin{tabular}{cccccc}
\hline
$x$ & $x_{BF}$ & $\langle x \rangle$ & $x_{med}$ & $68\%$ CL & $95\%$ CL \\
\hline \hline
~ & ~ & ~ & ~ & ~  & ~ \\
$\Omega_M$ & 0.283 & 0.278 & 0.277 & (0.260, 0.295) & (0.247, 0.308) \\
~ & ~ & ~ & ~ & ~  & ~ \\
$h$ & 0.704 & 0.705 & 0.705 & (0.694, 0.716) & (0.686, 0.728) \\
~ & ~ & ~ & ~ & ~  & ~ \\
$q_0$ & -0.78 & -0.79 & -0.78 & (-0.97, -0.61) & (-1.16, -0.52) \\
~ & ~ & ~ & ~ & ~  & ~ \\
$n$ & 1.44 & --- & --- & $\le 0.79$ & $\le 1.94$  \\
~ & ~ & ~ & ~ & ~  & ~ \\
$\log{\varepsilon}$ & 1.56 & 1.03 & 1.05 & (0.53, 1.48) & (0.24, 1.94) \\
~ & ~ & ~ & ~ & ~  & ~ \\
$\log{\tilde{\Lambda}_{eff}}$ & 2.86 & 2.68 & 2.67 & (1.51, 3.91) & (0.89, 4.42) \\
~ & ~ & ~ & ~ & ~  & ~ \\
$\sigma_{GRB}$ & 0.38 &  --- & --- & $\le 0.23$ & $\le 0.49$  \\
~ & ~ & ~ & ~ & ~  & ~ \\
\hline
\end{tabular}
\end{center}
\caption{Constraints on the St model parameters.}
\end{table}

\begin{displaymath}
\chi^2_{SNeIa}/d.o.f. = 1.03 \ , \ 
\chi^2_{GRB}/d.o.f. = 1.22 \ , \ 
\omega_M = 0.140 \ , 
\end{displaymath}
clearly show that the best fit St model is in very good
agreement with the data as can also be appreciated by looking at
Fig.\,\ref{fig: bfstplot}. Table II reports mean, median and 
confidence ranges for the individual parameters. An important
caveat is in order here for the parameters $n$ and $\sigma_{GRB}$. 
Indeed, the Markov chain tends to drift towards negative values of $n$
which are a priori excluded in order the St model to pass the theoretical 
constraints (\ref{eq: frconstr}). Therefore, the histogram of $n$ values
is clearly cut by this a priori assumption and, although formally mean 
and median values may be computed, they are not reliable. We thus give
only upper limits on $n$ because the data are unable to constrain this model 
parameter. A similar problem also takes place for the intrinsic scatter 
$\sigma_{GRB}$ which must be positive by definition. Nevertheless, the chain
drifts towards the border probably because of a degeneration
with $n$, with low $n$ preferring low scatter. We therefore only give upper 
limits on $\sigma_{GRB}$ too.

\begin{figure}
\centering \resizebox{8.5cm}{!}{\includegraphics{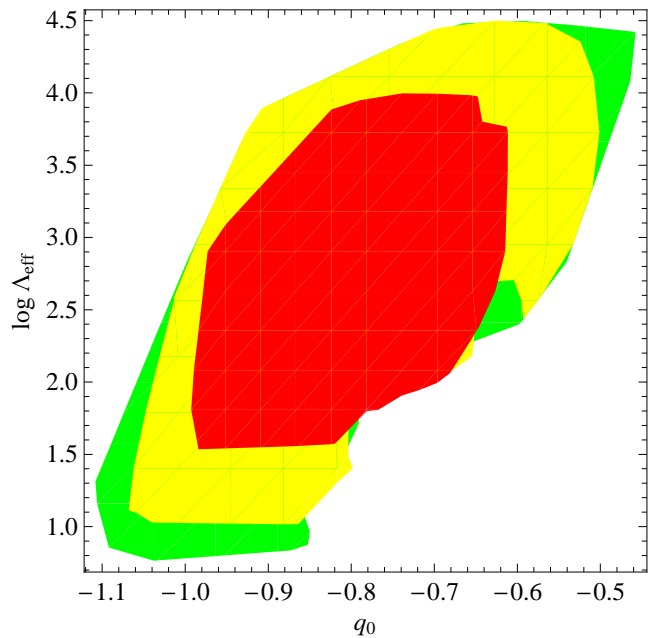}}
\caption{Likelihood contours in the $(q_0, \log{\tilde{\Lambda}_{eff}})$ plane.
Red, yellow and green shaded regions refer to the $68$, $95$, $99\%$ confidence regions 
respectively.}
\label{fig: qzlamcont}
\end{figure}

The comparison between Table I and II shows that the values of the three 
parameters $(\Omega_M, h, q_0)$ are consistent with each other. There
is actually some tension for the deceleration parameter $q_0$ with
the St model favouring a greater speed up, but the good overlap of 
the confidence ranges makes this difference statistically irrelevant. 
We do not discuss the comparison with previous results in the
literature but we refer to what we have said above for the HS model. As a remark,
however, we note that the agreement of the parameters $(\Omega_M, h, 
q_0)$ is not completely unexpected. Indeed, these quantities set 
the initial conditions (\ref{eq: incond}) so that this is only telling
us that the auxiliary functions $(y_H, y_R)$ are very close to 
each other for $z = 0$. Since both the HS and St models are designed
to fit the same present day data, it is not surprising that
they are quite similar today and lead to the same values of the
parameters which set the initial conditions.

It is, on the contrary, somewhat surprising that, notwithstanding
the initial larger acceleration, the transition redshift (estimated 
with the same procedure used above) is lower
than in the HS case, being\,:

\begin{displaymath}
\langle z_T \rangle = 0.44 \ \ , \ \ 
z_{T,med} = 0.43 \ \ ,
\end{displaymath}
\begin{displaymath}
{\rm 68\% \ CL \ :} \ \ (0.37, 0.52) \ \ , \ \ 
{\rm 95\% \ CL \ :} \ \ (0.33, 0.63) \ \ .
\end{displaymath}
Although formally in marginal disagreement at the $68\%$ level
with the results for the HS model, the two sets of constraints
are however not too different. In particular, the St $z_T$ value
remains in good agreement with the previous estimates in the literature.

Finally, even if we cannot put strong constraints on 
$(n, \log{\varepsilon}, \log{\tilde{\Lambda}_{eff}})$, we briefly
discuss their values.
First, we note that, since the term $R^2/(\varepsilon R_0)^2$ quickly starts
dominating over unity, at intermediate $z$, we get\,:
\begin{displaymath}
f(R) \sim -6 H_0^2 \tilde{\Lambda}_{eff} \left [ 
\left ( \frac{R}{\varepsilon R_0} \right )^{-2n} - 1 \right ] \ . 
\end{displaymath}
Let us suppose, for a moment, to drop the assumption $n > 0$. Indeed,
for $n < 0$, the above $f(R)$ is approximated by the sum of a (negative)
cosmological constant and a quadratic $R^2$ correction modulated by
the value of $\log{\varepsilon}$. Once again, we therefore recover a 
Lagrangian similar to the inflationary Lagrangian of Starobinsky, since the
negative cosmological constant soon becomes subdominant. This qualitative
discussion helps us understanding why the chain drifts towards negative
$n$ which have been excluded, in order to give Eq.(\ref{eq: frst}) the 
correct limit for $R \rightarrow \infty$. For $n > 0$, the code is forced
to look for an accelerating solution in this regime. Since, again, the 
term $R^2/(\varepsilon R_0)^2$ quickly overcomes the unity, the above
approximation of the St $f(R)$ still holds; we may easily note that 
the dominant correction is of the form $1/R^n$ so that the best fit $n = 1.44$
makes the model similar to the $1/R$ proposal firstly introduced as a fourth\,-\,order 
gravity motivated alternative to scalar field dark energy. As a final
remark, we look at Fig.\,\ref{fig: qzlamcont} which shows that $q_0$ is correlated
with the weighting parameter $\log{\tilde{\Lambda}_{eff}}$ which here plays the same 
role as $\log{c_1}$ for the HS model.  

\subsection{The effective dark energy EoS}

The impact of $f(R)$ on the expansion history can be alternatively
discussed by resorting to the effective dark energy equation of state 
(hereafter, EoS). Indeed, from the point of view of the background evolution, 
$f(R)$ models are equivalent to a cosmological scenario made out of  dust matter and an effective 
dark energy term with an EoS given by\,:
\begin{eqnarray}
1 + w_{eff}(z) & = & \left [ \frac{2}{3} \frac{d\ln{E(z)}}{d\ln{(1 + z)}} - 
\frac{\Omega_M (1 + z)^3}{E^{2}(z)} \right ] \nonumber \\
~ & \times & \left [ 1 - \frac{\Omega_M (1 + z)^3}{E^{2}(z)} \right ]^{-1} \ ,
\label{eq: eoseff}
\end{eqnarray}
so that the dark energy density parameter reads\,:
\begin{figure}
\centering \resizebox{8.5cm}{!}{\includegraphics{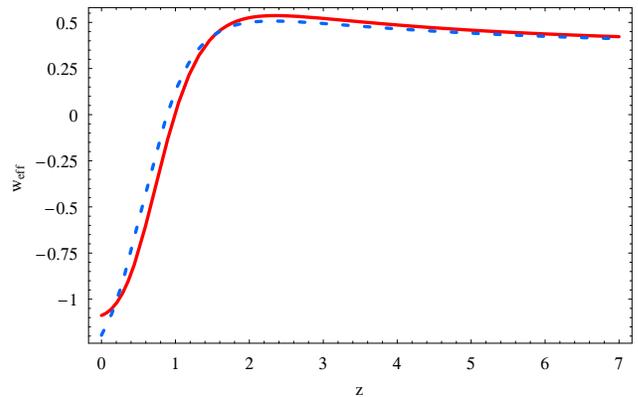}}
\caption{Effective EoS for the HS (red solid) and St (blue dashed) $f(R)$ theories
setting the model parameters to their best fit values.}
\label{fig: effeos}
\end{figure}
\begin{equation}
\Omega_{DE}(z) = \frac{1 - \Omega_M}{E^2(z)} 
\exp{\left [ 3 \int_{0}^{z}{\frac{1 + w_{eff}(z')}{1 + z'} dz'} \right ]} \ .
\label{eq: omdeeff}
\end{equation}
Fig.\,\ref{fig: effeos} shows the effective EoS for both the HS and St
models setting the parameters to their best fit values. It is impressive
to see how closely the two EoS are with $w_{eff}(z)$ being almost exactly the same
for the two models for $z > 2$. Note that here we have not plotted the full 
redshift range up to the last scattering $z \sim 1100$ to make easier to see
the details at low $z$. However, when $z \rightarrow 0$, we have checked that $w_{eff}(z)$ converges to 
$w_{eff} = 0$, i.e. the effective dark energy behaves as a matter term in the early universe
and scales with the redshift in the same way.
This is still an expected consequence of the constraints (\ref{eq: frconstr}) 
ensuring that $f(R)/R \rightarrow 0$ for $z \rightarrow \infty$. In such a limit,
we therefore recover a GR universe with only matter. Hence, we get $E^2(z \gg 1) \simeq
\Omega_M (1 + z)^3$ which, inserted into Eq.(\ref{eq: eoseff}) leads to $w_{eff}(z \gg 1)
= 0$ whatever is the $f(R)$ model considered. 

The HS and St effective EoS are different in the low $z$ regime. Indeed, for
the best fit models, we find\,:
\begin{displaymath}
w_{eff}(z = 0) = -1.09 \ \ , \ \
dw_{eff}/dz(z = 0) = 0.15 \ \ ,
\end{displaymath}
for the HS model and\,:
\begin{displaymath}
w_{eff}(z = 0) = -1.19 \ \ , \ \
dw_{eff}/dz(z = 0) = 0.69 \ \ ,
\end{displaymath}
for the St case. For both models, the present day values
of $w_{eff}$ and its redshift derivative disagree with the 
$\Lambda$CDM expected values, $w_{eff}(z = 0) = -1$ and 
$dw_{eff}/dz(z = 0) = 0$. Actually, we must consider the
Bayesian constraints obtained from evaluating these quantities
along the chain. For $w_{eff}(z = 0)$, we find\,:
\begin{displaymath}
\langle w_{eff}(z = 0) \rangle = -1.05 \ \ , \ \ 
[w_{eff}(z = 0)]_{med} = -1.04 \ \ ,
\end{displaymath}
\begin{displaymath}
{\rm 68\% \ CL \ :} \ \ (-1.10, -0.98) \ \ , \ \ 
{\rm 95\% \ CL \ :} \ \ (-1.30, -0.89) \ \ ,
\end{displaymath}
for the HS model, while for the St we get\,:
\begin{displaymath}
\langle w_{eff}(z = 0) \rangle = -1.19 \ \ , \ \ 
[w_{eff}(z = 0)]_{med} = -1.18 \ \ ,
\end{displaymath}
\begin{displaymath}
{\rm 68\% \ CL \ :} \ \ (-1.35, -1.03) \ \ , \ \ 
{\rm 95\% \ CL \ :} \ \ (-1.50, -0.95) \ \ .
\end{displaymath}
Defining $w_1 = dw_{eff}/dz(z = 0)$, we get\,:
\begin{displaymath}
\langle w_1 \rangle = 0.21 \ \ , \ \ 
w_{1,med} = 0.15 \ \ ,
\end{displaymath}
\begin{displaymath}
{\rm 68\% \ CL \ :} \ \ (-0.01, 0.35) \ \ , \ \ 
{\rm 95\% \ CL \ :} \ \ (-0.16, 1.14) \ \ ,
\end{displaymath}
for the HS model, while in the St case it is\,:
\begin{displaymath}
\langle w_1 \rangle = 0.83 \ \ , \ \ 
w_{1,med} = 0.69 \ \ ,
\end{displaymath}
\begin{displaymath}
{\rm 68\% \ CL \ :} \ \ (0.22, 1.20) \ \ , \ \ 
{\rm 95\% \ CL \ :} \ \ (-0.03, 4.05) \ \ .
\end{displaymath}
As it is apparent, an effective cosmological constant is 
reasonably consistent with these constraints. The present day
value of the EoS is indeed close to $-1$, while its first derivative, 
although being not null, is quite small and consitent with zero within
the $95\%$ confidence range. However, we stress
that neither the linear fit  $w(z) = w_0 + w_1 z$ nor
the widely used Chevallier\,-\,Polarski\,-\,Linder \cite{CPL}  
ansatz $w(z) = w_0 + w_a z/(1 + z)$ provide a good approximation over the redshift range probed.

As a final remark, let us look again at Fig.\,\ref{fig: effeos}.
For both $f(R)$ theories, the effective EoS takes a positive
value during the epoch of galaxy formation $(z \sim 2)$ and over a
wide redshift range in the matter dominated epoch. However, the 
effective dark energy density $\Omega_{DE}$ is very small during 
this period so that the deceleration parameter stays almost constant
to $q(z) \simeq 1$ for $z > 3$ as for a matter only universe. It is, 
therefore, likely that the departures from $w_{eff} = 0$ have essentially
no impact on the background when struture formation takes place.

\subsection{Local tests of gravity}

The $f(R)$ functional law for both the HS and St models has
been formulated in such a way to pass the constraints (\ref{eq: 
frconstr}). However,  this does not mean that there are no 
detectable deviations from the GR metric on Solar System
and galactic scales whatever the model parameters are. Indeed, 
the HS and St models correctly represent two different classes of 
$f(R)$ theories containing, as particular cases, models which are able to evade
the local tests of gravity. It is therefore interesting to explore
whether the cosmologically selected models belong to this subclass. A
key role will be played by the value of $|f_{R0}| = |df/dR|_{R = R_0}$. 
By evaluating this quantity along the chain, we get\,:
\begin{displaymath}
\langle \log{|f_{R0}|} \rangle = 3.87 \ \ , \ \ 
(\log{|f_{R0}|})_{med} = 3.81 \ \ ,
\end{displaymath}
\begin{displaymath}
{\rm 68\% \ CL \ :} \ \ (0.95, 6.34) \ \ , \ \ 
{\rm 95\% \ CL \ :} \ \ (-0.17, 10.01) \ \ 
\end{displaymath}
for the HS model and 
\begin{displaymath}
\langle \log{|f_{R0}|} \rangle = 0.16 \ \ , \ \ 
(\log{|f_{R0}|})_{med} = -0.03 \ \ ,
\end{displaymath}
\begin{displaymath}
{\rm 68\% \ CL \ :} \ \ (-0.20, 0.62) \ \ , \ \ 
{\rm 95\% \ CL \ :} \ \ (-0.32, 1.50) \ \ 
\end{displaymath}
for the St model. 

In order to see how these values compare with the constraints
from the local tests of gravity \cite{Will}, we follow \cite{HS07} who have
explicitly considered the case of the HS model. They have demonstrated
that, in order to be consistent with the constraints on the PPN
parameter $\gamma$, one has to impose the following condition\,:
\begin{equation}
f_{R0} < 74 (1.23 \times 10^6)^{n - 1} \left [ 
\frac{R_0}{m^2} \frac{\Omega_M h^2}{0.13} \right ]^{-(n + 1)} \ .
\label{eq: gammahs}
\end{equation}
Such a constraint thus depends on the value of the HS model parameters and 
could, in principle, be included to limit the volume of the parameter
space to be explored. However, we have preferred to let the MCMC code
be free of searching the full parameter space since (\ref{eq: gammahs}) is 
actually a quite weak prior. Indeed, for the best fit model, this reduces to
$\log{|f_{R0}|} < 13.4$, while the best fit value is $\log{|f_{R0}|} = 5.2$ so
that the test is easily passed. 

Unfortunately, we are unable to impose a similar constraint to 
the St model. To this aim, one should first solve the field equations for
a static spherically symmetric source matching the inner and outer solutions 
and taking care of a model for the Sun mass density profile. In \cite{HS07}, it
is argued that, for every $f(R)$ model, the PPN parameter $\gamma$ may always 
be expressed as,:
\begin{displaymath}
\gamma - 1 = - \frac{2 M_{eff}}{M_{eff} + M_{tot}}
\end{displaymath}
with $M_{tot}$ the total solar mass and 
\begin{displaymath}
M_{eff} = 4 \pi \int{(\rho - R/\kappa^2) r^2 dr}
\end{displaymath}
with $R(r)$ and $\rho(r)$ the scalar curvature of the local metric 
as function of the radial coordinate $r$ and $\rho(r)$ the mass density 
profile. In order to pass the Solar System constraints, $M_{eff}$ should be
as low as possible which can be accomplished by making $R(r)$ follow 
$\kappa^2 \rho(r)$ so that a chamaleon effect takes place. Since the St model
has been designed in such a way to develop such an effect, we are
confident that the Solar System constraints are fulfilled even if we cannot make
any quantitative estimate. 

Finally, let us consider the galactic scales where the following constraint
\begin{equation}
|f_{R0}| \le 2 \times 10^{-6} \left ( \frac{v_{max}}{300 \ {\rm km/s}} \right ) \ ,
\label{eq: frgalaxy}
\end{equation}
where $v_{max}$ is the  maximum rotation velocity of the stars, 
has been advocated in \cite{HS07} as a $f(R)$ indepedent condition to be
fulfilled in order to have no deviations of the gravitational potential
from the Keplerian $1/r$ profile on the galaxy scales. We remind that \cite{HS07}
look for a model describing the accelerated expansion of the Universe, but still
require a substantial dark matter content. However, the leading term $R^n$ of the 
Lagrangian of our best fit models induce a power law deviation from the Newtonian 
gravitational potential which was shown to describe the rotation curves of dwarf 
galaxies without the need of dark matter \cite{fogdm}. Therefore, it is not surprising 
that our cosmologically motivated constraints on $\log{|f_{R0}|}$ are definitively larger 
than the upper limit  $(\log{|f_{R0}|})_{max} \simeq -6$ for a Milky Way like galaxy. 
Although the result of the Lagrangian used by
\cite{fogdm} cannot be directly extrapolated to the HS and St models, it is however suggestive
that the violation of Eq.(\ref{eq: frgalaxy}) is an expected outcome of the region in the
parameter space preferred by the data. 

Rotation curves of spiral galaxies are an excellent probe of the gravitational
potential so that we can rely on them to judge whether any deviation from
the classical Newtonian result is detected. Indeed, the observed flatness of 
$v_c(r)$ in the outer regions, well far away from the edge of the visible disc,
clearly indicates that something unusual is taking place. If one assumes a priori 
that the potential remains Keplerian even on these scales, the only solution
is to invoke the presence of a halo made out of 
dark matter particles (which have not yet directly observed). 
In contrast, one can reject the dark halo hypothesis and 
thus interpret the flateness problem as an evidence for deviations from the Newtonian
potential. Should this interpretation be correct, one can try to fit the rotation curves
with a modified potential coming out of the low energy limit of a $f(R)$ theory, which 
has indeed been done with success \cite{fogdm}. As such, the fact that both the cosmologically
selected subclasses of the HS and St models violate the constraint (\ref{eq: frgalaxy})
should not be considered as a serious problem. Actually, giving off the dark halo in
favour of a modified potential could lead to a different problem. In such a case, indeed,
the matter density parameter should reduce to the baryons only one, i.e. $\Omega_M = 
\Omega_b \simeq 0.04$ in disagreement with the values in Tables I and II. A possible way out
could be to invoke massive neutrinos \cite{garry} or a dark matter component only present on 
cluster scales. 

As a final remark, however, it is also possible that the constraint (\ref{eq: frgalaxy}) is
overrestrictive or not valid at all. Indeed, Eq.(\ref{eq: frgalaxy}) has been
obtained assuming a spherical galaxy described by an NFW \cite{NFW} profile. Such a profile
is motivated by numerical simulations which assumes the validity of GR and hence a 
Newtonian gravitational potential. Therefore, testing the validity of GR based on a model
which yet assumes this is the case is somewhat a contradiction. Moreover, the derivation
in \cite{HS07} assumes that the galaxy is an isolated system, while galaxies actually
reside in groups or clusters. Should a long range fifth force be present, its effect must
be taken into account when deriving a constraint in Eq.(\ref{eq: frgalaxy}). As 
a consequence, an analytical computation which takes care of all these details is likely
to be impossible and one should resort to N\,-\,body simulations
in a $f(R)$ cosmological background, which are yet unavailable. 
Because of these problems, we warn the reader against considering the violation 
of (\ref{eq: frgalaxy}) by the HS and St $f(R)$ models as a definitive evidence that 
these theories should be ruled out.

\section{Conclusions}

As soon as the observational and conceptual problems related to the 
cosmological constant and other dark energy scenarios became
pressing, modification of the gravity sector of Einstein field equations
immediately appeared as an interesting alternative explanation of the 
observed cosmic speed up. Fourth\,-\,order gravity theories then appeared
as the most immediate generalizaton of Einstein GR since they just encoded
all the deviations into a single analytic function $f(R)$. As the problem
of acceleration was solved in this framework, a new problem came out, namely
how to choose a functional expression for $f(R)$ which is not only 
able to speed up the expansion (there were too many actually!), but also not to  
violate the local tests of gravity and turn off its effect in the early universe
where GR indeed correctly appears to work.

From a mathematical point of view, one has to look for a $f(R)$ expression
satisfying the constraints (\ref{eq: frconstr}) which is indeed what the 
HS and St models efficiently do, postulating that $f(R)$ is given by Eqs.(\ref{eq: frhs}) 
and (\ref{eq: frst}), respectively. Our aim here was then to test whether
these two well behaved models actually fit the data which
suggest the accelerated expansion of the Universe. To this end, we have 
considered the background evolution as probed by the Hubble diagram by using
both SNeIa and GRBs to cover the redshift range $(0.01, 6.6)$. As a first 
encouraging result, it turns out that both the HS and St models are in very 
good agreement with the data. Moreover, the predicted values of the matter
density parameter $\Omega_M$, the Hubble constant $h$, the deceleration parameter
$q_0$ and the transition redshift $z_T$ nicely compare with previous estimates
in the literature. As a further positive outcome, the cosmologically selected 
subclasses of the general HS and St parameterizations easily pass the constraints
on the $\gamma$ PPN parameter and reduce to GR in the early universe, as can
also be seen by noting that the corresponding effective EoS and density parameter
vanish at the redshift of the last scattering surface. Some unpleasant tension with the 
constraints on the deviation from the Newtonian potential on galaxy scales is
possible, but the way these constraints are derived put serious doubts on their
validity hence decreasing the significance of the problem. 

While this manuscript was near completion, a paper was posted on the arXiv 
\cite{MMA09} where the authors try to constrain the HS model parameters 
using the Union SNeIa sample, the Hubble expansion and age data \cite{SVJ05} and the 
Baryonic Acoustic Oscillations \cite{Eis05}. However, their results may not be
compared straightforwardly to ours in Table I. Notwithstanding the different
dataset used, the main difference relies in how they reparameterize the HS model. 
While we have directly used $(n, \log{c_1}, \log{c_2})$ as parameters, they 
prefer to use the set $(\tilde{\Omega}_M, n, f_{R0})$ with $\tilde{\Omega}_M$ the 
effective matter density parameter and $(c_1, c_2)$ estimated by\,:  
\begin{displaymath}
\frac{c_1}{c_2} \simeq 6 \frac{1 - \tilde{\Omega}_M}{\tilde{\Omega}_M} \ ,
\end{displaymath}
\begin{displaymath}
\frac{c_1}{c_2^2} = - \frac{f_{R0}}{n} \left ( \frac{12}{\tilde{\Omega}_M}
- 9 \right )^{n + 1} \ .
\end{displaymath}
Finally, they have limited a priori their exploration of the parameter space
to models with $|f_{R0}| \le 0.1$, while we have shown here that the best
fit is obtained for much larger values. In principle, we can solve the 
above relations to get $\tilde{\Omega}_M$ for our best fit model. But this is 
actually not possible since such relations have been obtained assuming 
that $|f_{R0}| \ll 1$ which is not true in our case. Such a condition was also imposed
in order to have $w_{eff}(z)$ not departing too much from a cosmological constant,
but we have shown here that this is an unnecessary assumption. Indeed, for both
the HS and St models $w_{eff}(z)$ is close to the $\Lambda$ EoS over a very 
limited range, but nevertheless we are able to fit the data since the impact of
the effective dark energy term becomes quickly subdominant during the matter 
dominated era no matter what its EoS is. 

As Fig.\,\ref{fig: effeos} shows, although the $f(R)$ functional
expression is different, both the HS and St best fit models have 
a very similar effective EoS and hence the predicted distance modulus 
is almost the same. Moreover, the values in Tables I and II tell us
that the SNeIa and GRBs Hubble diagram is unable to put strong constraints
on the model parameters. 

It is worth wondering how the situation can be
improved. As a first attempt, one can try to extend the redshift range
probed by relying on the use of the so-called distance priors \cite{WMAP5,San09}
which provide a useful set of distance-related quantities that summarize the
information contained in the CMBR anisotropy spectrum. Although widely
used as observational constraints in the recent literature \cite{dpused}, 
it is worth stressing that they are derived by assuming a fiducial $\Lambda$CDM
model. It is then argued that the mean and covariance matrix of the distance 
priors parameters do not change when the model space is enlarged, i.e. the choice 
of the  fiducial model does not impact the estimate of the priors. Moreover, one 
also assumes that the posterior probability from the CMBR anisotropies and galaxy 
power spectra is correctly described by the distance priors, i.e. no 
information is lost when dropping the full dataset in favour of the 
simplified one. While both these hypotheses have been verified for dark energy
models \cite{BW09}, such an exercise has never been performed for $f(R)$ theories. Therefore,
to be conservative, we have preferred not to use them in the present analysis.
One can, however, argue that including the distance priors can narrow down 
the constraints on some of the model parameters, but it is likely that the 
degeneracy between the HS and St models is not broken. Indeed, the distance priors
probe the universe mainly at the last scattering redshift $z_{LS} \simeq 1100$ thus
complementing SNeIa and GRBs which cover the range $(0, 7)$. However, in the 
early universe, both the HS and St models converge to GR so that they are likely 
to be too similar to be discriminated by probing this redshift regime.

Major improvements in the constraints and in the possibility to discriminate not
only between the HS and St models, but, more generally, among dark energy and 
$f(R)$ theories are expected from the analysis of the growth of perturbations. Indeed, 
even if one can tailor the $f(R)$ parameters in order to closely mimic the same 
expansion history of a given dark energy model, the evolution of 
the density perturbations can be rather 
different \cite{frstab,frPert,frSante}. In particular, this has a strong
impact on both the power spectrum and halo statistics \cite{frPS} and the 
weak lensing signals \cite{frWL}: it is thus possible to compare predictions
with data and severely constrain the viability of $f(R)$ theories. 

It is this combination of extended Hubble diagrams (made out of SNeIa and GRBs) and 
structure growth probes (such as galaxies power spectrum and cosmic shear) that
will finally tell us whether the observed cosmic speed up has been the first
evidence of a new fluid, as mysterious as fascinating, or of new physics in the 
gravity sector, as unexpected as challenging.  

\acknowledgments 

VFC warmly thanks S. Capozziello and M.G. Dainotti for their 
collaboration in getting the GRBs Hubble diagram and V. Salzano for
assistance with the Markov chains. VFC is supported by University 
of Torino and Regione Piemonte. The authors also ackwnoledge partial 
support from INFN grant PD51.

\end{document}